\documentclass[letterpaper,aps,twocolumn,showpacs,floats,prb]{revtex4}

\usepackage{amsmath}
\usepackage{graphicx,epsfig,dsfont}
\usepackage{amssymb}

\newcommand{\nn}{\nonumber}

\begin{document}

\title{Scattering of high-energy magnons off a magnetic skyrmion}

\author{Sarah Schroeter}
\author{Markus Garst}
\affiliation{Institut f\"ur Theoretische Physik, Universit\"at zu K\"oln,
Z\"ulpicher Str. 77a, 50937 K\"oln, Germany
}

\begin{abstract}
We discuss the scattering of high-energy magnons off a single magnetic skyrmion within the field-polarized ground state of a two-dimensional chiral magnet. For wavevectors larger than the inverse skyrmion radius, $k r_s \gg 1$, the magnon scattering is dominated by an emerging magnetic field whose flux density is essentially determined by the topological charge density of the skyrmion texture. This leads to skew and rainbow scattering characterized by an asymmetric and oscillating differential cross section.
We demonstrate that the transversal momentum transfer to the skyrmion 
is universal due to the quantization of the total emerging flux
while the longitudinal momentum transfer is negligible in the high-energy limit. This results in a magnon-driven skyrmion motion approximately antiparallel to the incoming magnon current and a universal relation between current and skyrmion-velocity.
\end{abstract}

\date{\today}

\maketitle

\section{Introduction}

The experimental discovery of skyrmions in chiral magnets \cite{Muehlbauer2009,Muenzer2010,Yu2010,Yu2011,Adams2011,Seki2012,Adams2012} and in 
magnetic monolayers \cite{Heinze2011,Romming2013,Bergmann2014} has triggered 
an increasing interest in the interaction of spin currents with topological magnetic textures.\cite{Neubauer2009,Lee2009,Jonietz2010,Everschor2011,Everschor2012,Yu2012,Schulz2012,Iwasaki2013,Lin2013-1,Lin2013-2,Sampaio2013,Nagaosa2013,Kong2013,Lin2013-3,Mochizuki2014,Lin2014,Iwasaki2014,Kovalev2014,Schuette2014,Franz2014,Schuette2014-2,Mueller2015}
It has been demonstrated\cite{Jonietz2010,Yu2012} that skyrmions can be manipulated by ultralow electronic current densities of $10^6$ A/m$^2$, which is five orders of magnitudes smaller than in conventional spintronic applications using domain walls. The adiabatic spin-alignment of electrons moving across a skyrmion texture results in an emergent electrodynamics implying a topological \cite{Neubauer2009,Lee2009,Franz2014} as well as a skyrmion-flow Hall effect.\cite{Schulz2012} 
In insulators, the interplay of thermal magnon currents and skyrmions is marked by a topological magnon Hall effect and a magnon-driven skyrmion motion.\cite{Kong2013,Lin2013-3,Mochizuki2014} 
The topological nature of the magnetic skyrmions is responsible for a peculiar dynamics\cite{Petrova2011,Zang2011,Mochizuki2012,Onose2012,Schwarze2014} that is also at the origin of these novel spintronic and caloritronic phenomena, which are at the focus of the fledgling field of {\it skyrmionics}.\cite{Nagaosa2013}

\begin{figure}
\centering
\includegraphics[width=0.8\columnwidth]{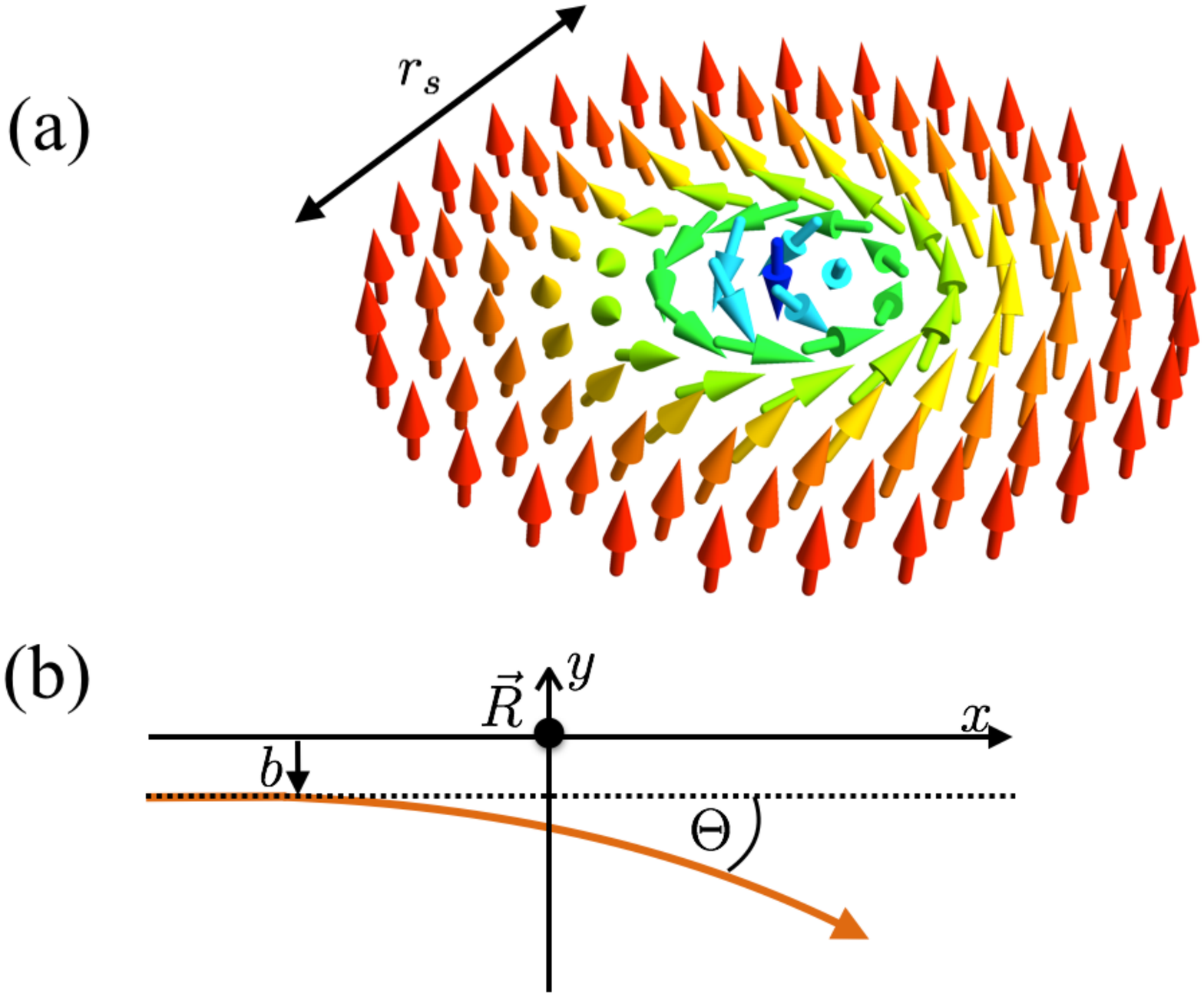}
\caption{(a) A  chiral magnetic skyrmion texture of linear size $r_s$. (b) Illustration of a classical magnon trajectory within the $x$-$y$ plane scattering off a skyrmion positioned at $\vec R$ with impact parameter $b$ and classical deflection angle $\Theta$. 
}
\label{fig:Illustration}
\end{figure}

In two spatial dimensions, skyrmions are identified by the topological charge density 
\begin{align}
\rho_{\rm top} =  \frac{1}{4\pi} \hat n (\partial_x \hat n \times \partial_y \hat n),
\end{align}
where $\hat n$ is the orientation of the magnetization vector. For a magnetization homogeneously polarized at the boundary, the spatial integral $\int d^2 {\bf r} \rho_{\rm top} = W$ is quantized, $W \in \mathds{Z}$, and thus allows to count skyrmions within the sample. In turn, a finite winding number $W$ translates to a gyrocoupling vector $\vec G$ in the Thiele equation of motion of the skyrmion,\cite{Thiele1973} and the resulting gyrotropic spin-Magnus force governs its dynamics.\cite{Stone1996} 
As a consequence, in the presence of an applied electronic spin current,
the skyrmions will acquire  a velocity\cite{Everschor2011,Schulz2012,Everschor2012} 
that remains finite in the limit of adiabatic spin-transfer torques and small Gilbert damping $\alpha$, giving rise to a universal current-velocity relation.\cite{Iwasaki2013}

In order to address the interaction of magnon currents with magnetic textures, 
a corresponding adiabatic approximation has been recently invoked on 
the level of the Landau-Lifshitz-Gilbert equation by Kovalev and Tserkovnyak.\cite{Kovalev2012} 
This approximation has been used in Refs.~\onlinecite{Kong2013,Lin2013-3} to derive an effective Thiele equation of motion for the skyrmion coordinate $\vec R$ in the presence of a magnon current density $\vec J$,
\begin{align} \label{AdiabaticThiele}
\vec G \times \dot{\vec R} = - \vec G \times \vec v_{\rm eff} + \beta \vec v_{\rm eff} + \dots,
\end{align}
with $\beta = 0$ in the adiabatic limit.
The effective velocity $\vec v_{\rm eff} = g \mu_B \vec J/(\hbar m_0)$ is related to the current density via the $g$-factor $g$, the Bohr magneton $\mu_B > 0$ and the local magnetization $m_0$. The gyrocoupling vector is given by $\vec G = - 4\pi  \hat z \hbar m_0/(g\mu_B) $ with units of spin density corresponding to a flux of $-2\pi\hbar$ per area of a spin-$\frac{1}{2}$ in a two-dimensional system with the unit normal vector $\hat z$. The dots in Eq.~\eqref{AdiabaticThiele} represent further terms omitted for the purpose of the following discussion, that is, in particular, a damping force proportional to the Gilbert constant $\alpha$.
Neglecting these additional terms, Eq.~\eqref{AdiabaticThiele} predicts for $\beta = 0$, similar to the skyrmion-driven motion by electronic currents, a universal current-velocity relation $\dot{\vec R} = - \vec v_{\rm eff} = - g \mu_B \vec J/(\hbar m_0)$ with a skyrmion velocity that is antiparallel to $\vec J$. Consequently, a magnon current generated by a thermal gradient will induce a skyrmion motion towards the hot region of the sample, which was indeed observed numerically.\cite{Kong2013,Lin2013-3,Iwasaki2014}
Mochizuki {\it et al.}\cite{Mochizuki2014} also used Eq.~\eqref{AdiabaticThiele} with $\beta=0$ to account for the experimental observation of a thermally induced rotation of a skyrmion crystal.

However, the question arises as to when the adiabatic limit of Eq.~\eqref{AdiabaticThiele} is actually applicable and under what conditions. The validity regime of the adiabatic approximation for magnon-driven motion of magnetic textures has not been explicitly discussed in Ref.~\onlinecite{Kovalev2012}. In fact, in order to account quantitatively for 
their numerical experiment Lin {\it et al.}\cite{Lin2013-3} introduced  the $\beta$ parameter in Eq.~\eqref{AdiabaticThiele} on phenomenological grounds calling it a measure for non-adiabaticity. Subsequently, Kovalev\cite{Kovalev2014} argued that a finite $\beta$ parameter arises due to dissipative processes. 

In contrast, we have recently shown by considering the magnon-skyrmion scattering problem\cite{Schuette2014} that a monochromatic magnon current with energy $\varepsilon$ will give rise to 
a {\it reactive} momentum-transfer force in the Thiele equation which reads in linear response
\begin{align} \label{MomentumTransferThiele}
\vec G \times \dot{\vec R} =  k \sigma_\perp(\varepsilon) (\hat z \times \vec J_\varepsilon) + k \sigma_\parallel(\varepsilon) \vec J_\varepsilon  + \dots,
\end{align}
where the magnon dispersion is $\varepsilon = \varepsilon_{\rm gap} + (\hbar k)^2/(2M_{\rm mag})$ with the magnon gap $\varepsilon_{\rm gap}$ and the magnon mass $M_{\rm mag}$. 
This force on the right-hand side of Eq.~\eqref{MomentumTransferThiele} is determined by the two-dimensional transport scattering cross sections
\begin{align} \label{TrCS}
\left(\begin{array}{c} \sigma_\parallel(\varepsilon) \\ \sigma_\perp(\varepsilon) 
\end{array} \right) = \int_{-\pi}^\pi d\chi 
\left(\begin{array}{c} 1 - \cos \chi \\ -\sin \chi \end{array} \right)
\frac{d \sigma}{d\chi}
\end{align}
where $\frac{d \sigma}{d\chi}$ is the energy-dependent differential scattering cross section of the skyrmion.
In the limit of low energies $k r_s \ll 1$, where $r_s$ is the skyrmion radius, $s$-wave scattering is found to dominate 
so that $\sigma_\perp(\varepsilon) \to 0$ and, as shown in Ref.~\onlinecite{Schuette2014}, the force becomes longitudinal to $\vec J_\varepsilon$. This, in turn, implies a skyrmion motion approximately perpendicular to the magnon current, $\dot{\vec R} \to \frac{k \sigma_\parallel(\varepsilon)}{|\vec G|}\hat z \times \vec J_\varepsilon$, thus maximally violating the predictions of the adiabatic limit of Eq.~\eqref{AdiabaticThiele}. This implies that Eq.~\eqref{AdiabaticThiele} is not valid for low-energy magnons whose wavevector is comparable or smaller than the inverse size of the  texture.

It is one of the aims of this work to demonstrate explicitly that in the high-energy limit, $k r_s \gg 1$, on the other hand, the momentum-transfer force of Eq.~\eqref{MomentumTransferThiele} due to a monochromatic magnon wave indeed reduces to the form of Eq.~\eqref{AdiabaticThiele}.
The effective velocity in this case, however, is to be identified with $\vec v_{\rm eff} = |A|^2 \hbar \vec{k}/M_{\rm mag}$ where $A$ is the amplitude of the incoming magnon wave.
In the high-energy limit the magnon-skyrmion interaction is dominated by a scattering vector potential, i.e., an emerging orbital magnetic field whose  flux is quantized and related to the skyrmion topology. As a result, the transversal momentum transfer assumes a universal value in the high-energy limit $k \sigma_\perp(\varepsilon) \to 4\pi$  as anticipated in Ref.~\onlinecite{Mochizuki2014}. Moreover, the longitudinal momentum transfer yields a reactive contribution, $\beta_\varepsilon$, to the $\beta$ parameter that, in this limit, is determined by the square of the classical deflection function $\Theta(b)$ integrated over the impact parameter $b$, see Fig.~\ref{fig:Illustration}(b),
\begin{align} \label{BetaReactive}
\beta_\varepsilon = \frac{|G|}{8\pi} k \int_{-\infty}^{\infty}db\, (\Theta(b))^2.
\end{align}
As the scattering is in forward direction at high energies, $\Theta(b) \sim 1/k$, the parameter vanishes as $\beta_\varepsilon \propto 1/k$ so that it is indeed small for large $k r_s \gg 1$.

The outline of the paper is as follows. In section \ref{sec:Model} we shortly review the definition of the magnon-skyrmion scattering problem and some of the main results of Ref.~\onlinecite{Schuette2014}. In section \ref{sec:HighEnergyScatt} we examine the scattering properties of high-energy magnons including the skew and rainbow effects, the total and transport scattering cross sections, and the magnon pressure on the skyrmion leading to Eq.~\eqref{AdiabaticThiele}. We finish with a short discussion in section \ref{sec:discussion}. 

\section{Skyrmionic soliton and its spin-wave excitations}
\label{sec:Model}

This section closely follows Ref.~\onlinecite{Schuette2014} and reviews the magnon-skyrmion scattering problem in a two-dimensional chiral magnet.
We start with the standard model for a cubic chiral magnet restricted to a two-dimensional plane that is described by the energy functional\cite{Bak80,Nakanishi80}
\begin{align} \label{NLsM1}
\mathcal{E} = \frac{\rho_s}{2} \Big[ (\partial_\alpha \hat n_j)^2 
+ 2 Q \epsilon_{i\alpha j} \hat n_i \partial_\alpha \hat n_j - 2 \kappa^2 \hat n \hat B\Big]
\end{align}
with spatial index $\alpha \in \{1,2\} = \{x,y\}$ and $i,j \in\{1,2,3\}$, $\epsilon_{i\alpha j}$ is the totally antisymmetric tensor with $\epsilon_{123} = 1$, and $\rho_s$ is the stiffness. 
The two length scales are given by the wavevectors $Q$ and $\kappa$. The former determines the strength of the spin-orbit Dzyaloshinskii-Moriya interaction, that we chose to be positive, $Q>0$. The latter, $\kappa>0$, measures the strength of the applied magnetic field, that is applied perpendicular to the two-dimensional plane, $\hat B = \hat z$. We neglect cubic anisotropies, dipolar interactions as well as magnetic anisotropies for simplicity. The latter can be easily included resulting in an additional length scale. 

\subsection{Skyrmionic saddle-point solution}

The theory \eqref{NLsM1} possesses a topological soliton solution, i.e., a skyrmion, as first pointed out by Bogdanov and Hubert.\cite{Boganov1994,Roessler2006} With the standard parametrization of the unit vector $\hat n_s^T = (\sin \theta \cos \varphi, \sin \theta \sin \varphi, \cos \theta)$, the skyrmion obeys 
\begin{align}
\theta = \theta(\rho),\quad \varphi = \chi + \frac{\pi}{2},
\end{align}
where $\rho$ and $\chi$ are polar coordinates of the two-dimensional spatial vector ${\bf r} = \rho (\cos \chi, \sin \chi)$. The polar angle $\theta$ obeys the differential equation
\begin{align}
\theta'' + \frac{\theta'}{\rho} - \frac{
\sin \theta \cos \theta}{\rho^2} + \frac{2 Q \sin^2 \theta}{\rho} - \kappa^2 \sin \theta = 0, 
\end{align}
with the boundary conditions $\theta(0) = \pi$ and $\lim_{\rho \to \infty} \theta(\rho) = 0$. 
At large distances $\rho\kappa \gg 1$, the polar angle obeys the asymptotics $\theta(\rho) \sim e^{-\kappa \rho}/\sqrt{\rho}$, which identifies $\kappa$ as the inverse skyrmion radius. The resulting skyrmion texture is illustrated in Fig.~\ref{fig:Illustration}(a).
The associated topological charge density
\begin{align} \label{TopCharge}
\rho^s_{\rm top} = \frac{1}{4\pi} \hat n_s (\partial_x \hat n_s \times \partial_y \hat n_s) = \frac{1}{4\pi} \frac{\theta' \sin \theta}{\rho}
\end{align}
integrates to $\int d^2{\bf r} \rho^s_{\rm top} = -1$ identifying the solution as a skyrmion. The skyrmion radius $r_s$ can be defined with the help of the area 
$\int d^2 r (1-\hat n_z)/2 = \pi r_s^2$, and it is found to approximately obey $r_s \sim 1/\kappa^2$.

The skyrmion is a large-amplitude excitation of the fully polarized ground state as long as its energy is positive, which is the case for $\kappa > \kappa_{\rm cr}$ where $\kappa_{\rm cr}^2 \approx 0.8 Q^2$, which is the regime we focus on. For smaller values of $\kappa$, skyrmions proliferate resulting in the formation of a skyrmion crystal ground state.

\subsection{Magnon-skyrmion scattering problem}

\subsubsection{Magnon wavefunction}

The magnons correspond to spin-wave excitations around the skyrmion solution $\hat n_s$ that can be analyzed in the spirit of previous work by Ivanov and collaborators. \cite{Ivanov1995,Ivanov1998,Sheka2001,Sheka2004}
We introduce the local orthogonal frame $\hat e_i \hat e_j = \delta_{ij}$ with $\hat e_1 \times \hat e_2 = \hat e_3$, where $\hat e_3({\bf r}) = \hat n_s({\bf r})$ tracks the skyrmion profile. For the two orthogonal vectors we use $\hat e^T_1 = (-\sin \varphi, \cos \varphi,0)$ and $\hat e^T_2 = (-\cos \theta \cos \varphi, \sin \theta \sin \varphi, \cos \theta)$. The excitations are parametrized in the standard fashion 
\begin{align}
\hat n = \hat e_3 \sqrt{1-2 |\psi|^2} +\hat e_+ \psi + \hat e_- \psi^*,
\end{align}
where $\psi$ is the magnon wavefunction and $\hat e_\pm = \frac{1}{\sqrt{2}} (\hat e_1 \pm i \hat e_2)$.
For large distances, $\rho \gg r_s$, this parametrization assumes the form 
\begin{align} \label{FrameAsympt}
\hat n \approx \hat z \sqrt{1-2 |\psi|^2} + \Big(\frac{1}{\sqrt{2}}(\hat x + i \hat y) (-e^{-i \chi} \psi) + c.c.\Big).
\end{align}
It is important to note that the local frame $\hat e_i$ corresponds to a rotating frame even at large distances 
reflected in the phase factor $-e^{-i \chi}$ in the second term. For the discussion of magnon scattering, it will be convenient to introduce a wavefunction $\psi_{\rm lab}$ with respect to a frame that reduces to the laboratory frame at large distances, that is simply obtained by the gauge transformation
\begin{align} \label{GaugeTransformation}
\psi_{\rm lab}({\bf r}, t) = -e^{-i \chi} \psi({\bf r}, t).
\end{align}

\subsubsection{Magnon Hamiltonian}

In order to derive an effective Hamiltonian for $\psi$, we consider the 
Landau-Lifshitz equation 
\begin{align} \label{LL}
\partial_t \hat n = - \gamma \hat n \times \vec B_{\rm eff},
\end{align}
with $\gamma = g \mu_B/\hbar$, where the effective magnetic field $\vec B_{\rm eff}({\bf r},t) = -\frac{1}{m_0} \frac{\delta E}{\delta \hat n({\bf r},t)}$ is determined by the functional derivative of the integrated energy density $E = \int dt d{\bf r} \mathcal{E}$.
Expanding \eqref{LL} in lowest order in $\psi$, one finds that the spinor $\vec \Psi^T = (\psi, \psi^*)$ is governed by a bosonic Bogoliubov-deGennes (BdG) equation
\begin{align} \label{BdG}
i\hbar \tau^z \partial_t \vec \Psi = \mathcal{H} \vec \Psi,
\end{align}
with the Hamiltonian
\begin{align}
\mathcal{H} = 
\frac{\hbar^2 (- i \mathds{1} \vec \nabla - \tau^z \vec a)^2}{2 M_{\rm mag}} +  \mathds{1} \mathcal{V}_0 + \tau^x \mathcal{V}_x,
\end{align}
where $\vec \nabla^T = (\partial_x, \partial_y)$, and $\tau^x$ and $\tau^z$ are Pauli matrices. The potentials are given by 
\begin{align}
\mathcal{V}_0(\rho) = \frac{\varepsilon_{\rm gap}}{\kappa^2}&\Big(
-\frac{\sin^2 \theta}{2\rho^2} - \frac{Q \sin(2 \theta)}{2\rho} ,
\\\nn
&- Q^2 \sin^2 \theta 
+\kappa^2 \cos\theta - Q \theta' - \frac{\theta'^2}{2}\Big)
\\
\mathcal{V}_x(\rho) = \frac{\varepsilon_{\rm gap}}{\kappa^2}&\Big(
\frac{\sin^2 \theta}{2\rho^2} + \frac{Q \sin(2 \theta)}{2\rho}  - Q \theta' - \frac{\theta'^2}{2}
\Big).
\end{align}
The magnon energy gap is defined by 
\begin{align}
\varepsilon_{\rm gap} = \frac{g \mu_B \rho_s \kappa^2}{m_0} = \frac{\hbar^2 \kappa^2}{2 M_{\rm mag}},
\end{align}
which also identifies the magnon mass $M_{\rm mag}$.
The vector potential reads $\vec a = a^\chi(\rho) \hat \chi$ with $\hat \chi^T = (-\sin\chi,\cos \chi)$ and 
\begin{align} \label{VectorPot}
a^\chi &=   \frac{\cos \theta}{\rho} - Q \sin \theta.
\end{align}
It obeys the Coulomb gauge $\nabla \vec a = 0$.
The polar angle in all potentials is the soliton solution, $\theta = \theta(\rho)$, and depends on the distance $\rho$.

\subsubsection{Effective magnetic flux}

Far away from the skyrmion the Hamiltonian simplifies $\mathcal{H} \to \mathcal{H}_0$ for $\rho \to \infty$ with 
\begin{align}
\mathcal{H}_0 = 
\frac{\hbar^2 (- i \mathds{1} \vec \nabla - \tau^z \frac{1}{\rho}  \hat \chi)^2}{2 M_{\rm mag}} +  \mathds{1} \varepsilon_{\rm gap}.
\end{align}
The remaining vector potential is attributed to the choice of the rotating orthogonal frame in the definition of the magnon wavefunction, see Eq.~\eqref{FrameAsympt}. It can be easily eliminated by the gauge transformation \eqref{GaugeTransformation},
\begin{align}
\vec \Psi \to \vec \Psi_{\rm lab} &= e^{-i \tau^z (\chi+\pi)} \vec \Psi 
\\ \label{VectorPotLab}
a^\chi \to a^\chi_{\rm lab} &= a^\chi - \frac{1}{\rho} =  \frac{\cos \theta - 1}{\rho} - Q \sin \theta.
\end{align}
With respect to this laboratory orthogonal frame, the vector scattering potential $\vec a_{\rm lab} = a^\chi_{\rm lab} \hat \chi$ vanishes exponentially for large distances, $\rho \gg r_s$. 

\begin{figure}
\centering
\includegraphics[width=0.8\columnwidth]{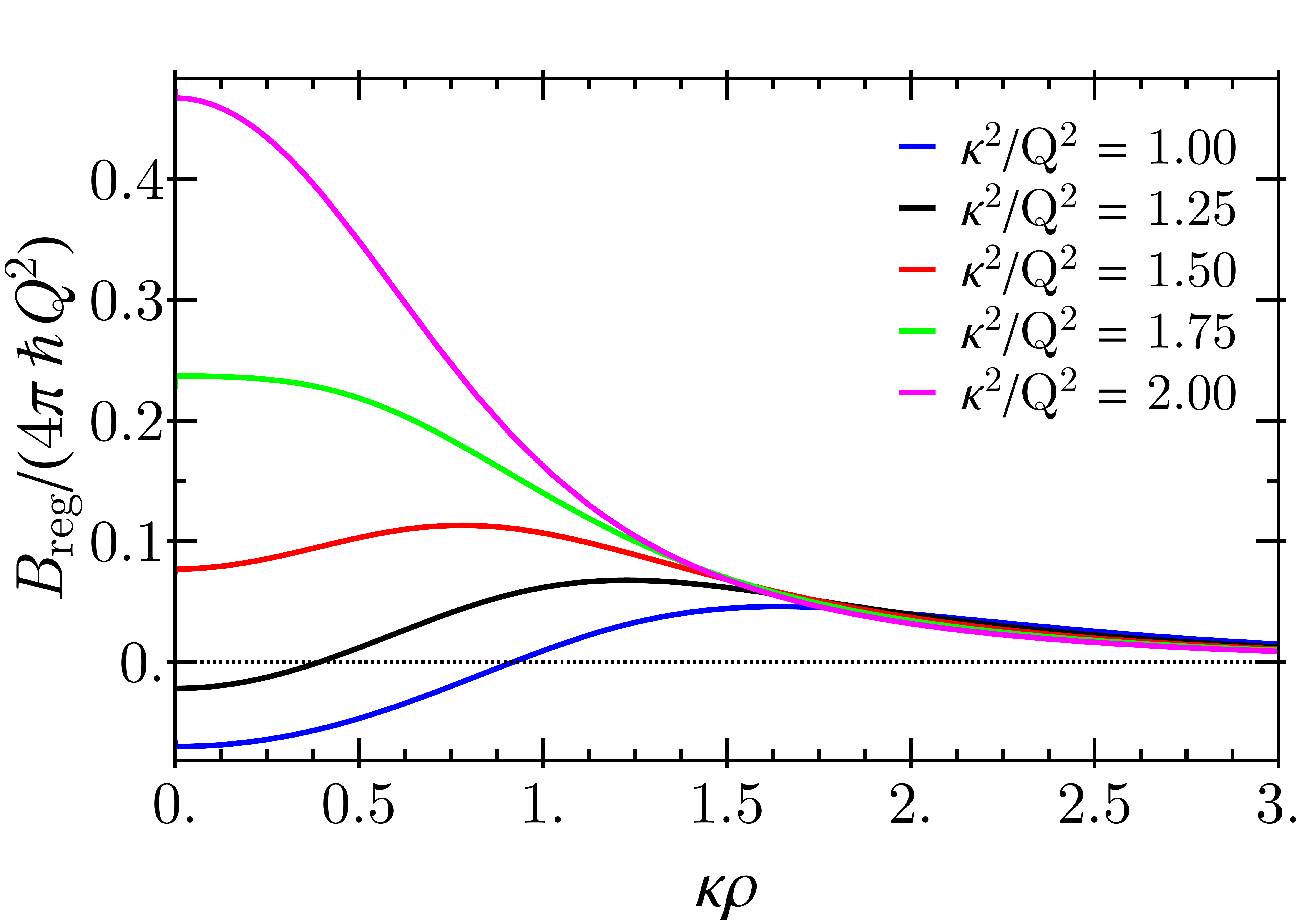}
\caption{Regular part of the effective magnetic flux density \eqref{Breg} for various values of $\kappa^2/Q^2$.
For lower values of $\kappa^2/Q^2$ the flux density close to the skyrmion center is suppressed and even becomes negative for $\kappa^2/Q^2 \lesssim 1.3$. 
As a result, the effective local Lorentz force evaluated along a classical magnon trajectory with $b=0$
changes sign resulting in a suppression of the deflection angle.
}
\label{fig:Flux}
\end{figure}

The associated flux $\vec{\mathcal{B}} = \nabla \times (\hbar  \vec a_{\rm lab})= \mathcal{B} \hat z$ will play an important role in the following discussion, where $\mathcal{B}({\bf r}) = \frac{\hbar}{\rho}\partial_\rho (\rho a^\chi_{\rm lab}(\rho))$.
According to Stokes'  theorem the total flux $\int d^2 {\bf r} \mathcal{B}({\bf r}) =0$ vanishes as $\vec a_{\rm lab}$ is exponentially confined to the skyrmion radius. However, there is an interesting spatial flux distribution,
\begin{align}
\mathcal{B}({\bf r}) &= - 4\pi\hbar \delta({\bf r}) + \mathcal{B}_{\rm reg}(|{\bf r}|),
\\ \label{Breg}
\mathcal{B}_{\rm reg}(\rho) &= 4 \pi \hbar\left(- \rho^s_{\rm top} - \frac{Q}{4\pi \rho} \partial_\rho(\rho \sin\theta) \right).
\end{align}
Since for small distances $a^\chi_{\rm lab}(\rho) \to - 2/\rho$, there is a singular flux contribution at the skyrmion origin with quantized strength $- 4\pi\hbar$. As it is quantized, this singular flux will not contribute to the magnon scattering. The regular part of the effective magnetic flux, $\mathcal{B}_{\rm reg}$, only depends on the radius $\rho$ and is spatially confined to the skyrmion area. Its spatial distribution can be related with the help of Eq.~\eqref{TopCharge} to the topological charge density $\rho^s_{\rm top}$ of the skyrmion in addition to a term proportional to $Q$. While $- \rho^s_{\rm top}$ is always positive, the latter term can also be negative so that $\mathcal{B}_{\rm reg}$  as a function of distance $\rho$ even changes sign for lower values of $\kappa^2$, see Fig.~\ref{fig:Flux}.  The spatial integral over the second term of Eq.~\eqref{Breg} however vanishes so that the total regular flux 
$\int d^2 {\bf r} \mathcal{B}_{\rm reg}(\rho) = - 4 \pi \hbar \int d^2 {\bf r} \rho^s_{\rm top} = 4\pi \hbar$ is quantized and determined by the topological charge of the skyrmion.\cite{Ivanov2005,Mochizuki2014}

\subsection{Magnon spectrum}

\begin{figure}
\centering
\includegraphics[width=0.8\columnwidth]{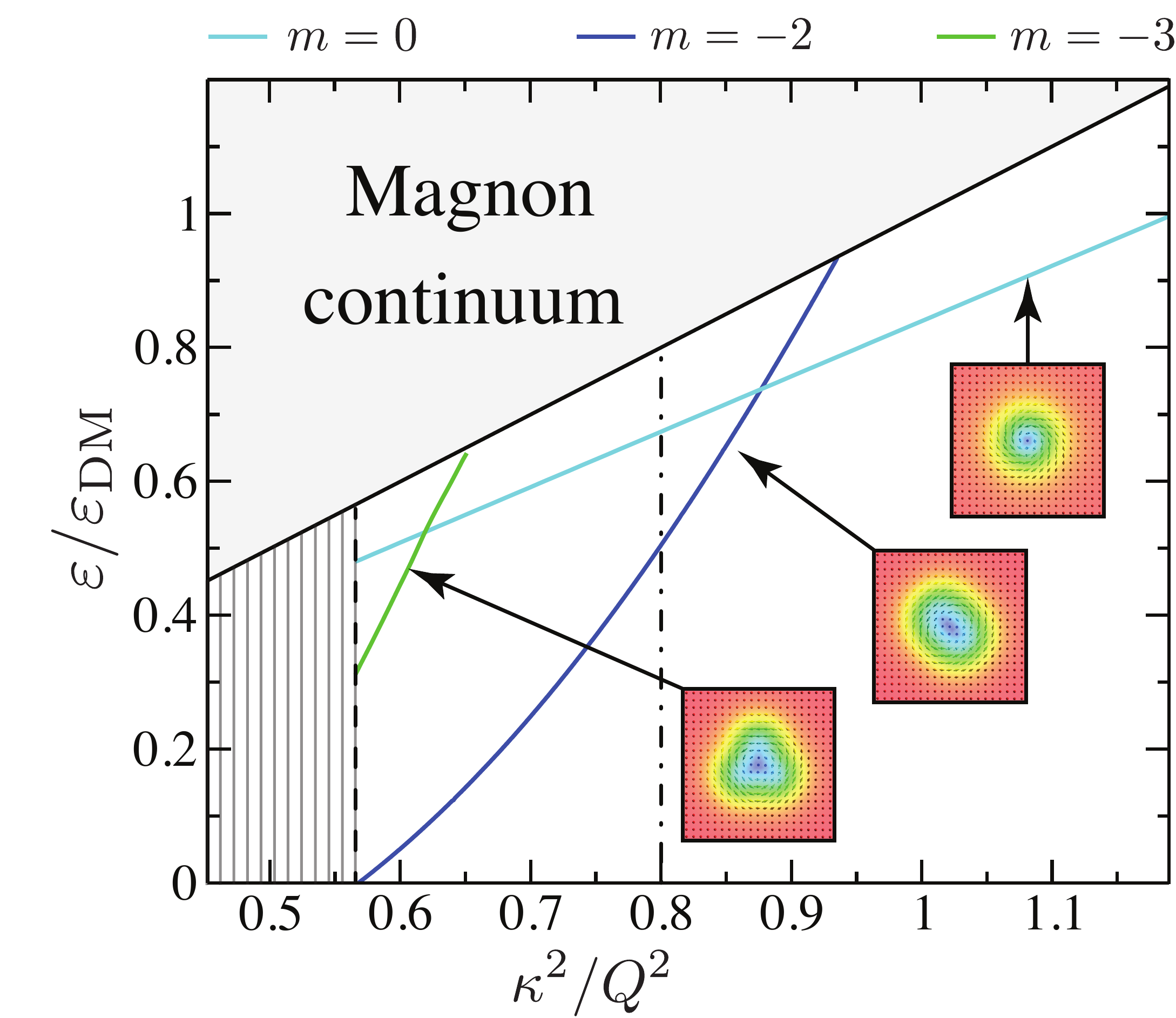}
\caption{Magnon spectrum in the presence of a single skyrmion excitation as a function of $\kappa^2/Q^2$ measuring the strength of the magnetic field.\cite{Schuette2014} The magnon gap $\varepsilon_{\rm gap} = \varepsilon_{\rm DM} \kappa^2/Q^2$ increases linearly with the field (black solid line). The field-polarized state becomes unstable at $\kappa^2_{\rm cr} \approx 0.8 Q^2$ (dashed-dotted line) while the theory \eqref{BdG} becomes locally unstable at $\kappa^2_{\rm bimeron} \approx 0.56 Q^2$. Apart from the zero mode (not shown), there exist three subgap modes with angular momentum $m=0,-2,-3$.
}
\label{fig:spectrum}
\end{figure}

In order to solve Eq.~\eqref{BdG} for the magnon eigenvalues and eigenfunctions, one uses the angular momentum basis $\vec \Psi({\bf r},t) = e^{- i \varepsilon t/\hbar + i m\chi} \vec \eta_m(\rho)$ with positive energy $\varepsilon\geq0$. The angular momentum $\hbar m$ turns out to be a good quantum number and the wave equation \eqref{BdG} reduces to a radial eigenvalue problem for $\vec \eta_m(\rho)$
that can be solved with the help of the shooting method.\cite{Schuette2014} In order to obtain positive expectation values of the Hamiltonian, one has to look for eigenfunctions with a positive norm,
\begin{align}
\int_0^\infty d\rho \rho\, \vec \eta^\dagger_m(\rho) \tau^z \vec \eta_m(\rho) > 0.
\end{align}

The resulting spectrum is shown in Fig.~\ref{fig:spectrum} as a function of the parameter $\kappa^2/Q^2$ that measures the strength of the magnetic field. The magnon continuum with the scattering states are confined to energies larger than the magnon gap $\varepsilon_{\rm gap} \propto \kappa^2$, which increases linearly with the field (black solid line). In the field range shown, there are three subgap states that correspond to bound magnon-skyrmion modes. While the breathing mode with angular momentum $m=0$ exists over the full field range, a quadrupolar mode with $m=-2$ emerges for lower fields just before the field-polarized state becomes globally unstable at $\kappa^2_{cr} \approx 0.8 Q^2$ (dashed-dotted line). The eigenenergy of the latter finally vanishes at $\kappa^2_{\rm bimeron} \approx 0.56 Q^2$, indicating a local instability of the theory with respect to quadrupolar deformations of the skyrmion, i.e. the formation of a bimeron.\cite{Ezawa2011-2} 
Furthermore, a sextupolar mode with $m=-3$ only exists within the metastable regime. 
The corresponding eigenfunctions of these modes do not possess any nodes. 

Apart from the modes shown in Fig.~\ref{fig:spectrum}, the spectrum of $\mathcal{H}$ also contains a zero mode with angular momentum $m=-1$ given by 
\begin{align} \label{ZeroMode}
\vec \eta^{\,\rm zm}_{-1} = \frac{1}{\sqrt{8}} \left(\begin{array}{c}
\frac{\sin \theta}{\rho} - \theta' \\ \frac{\sin \theta}{\rho} + \theta'
\end{array}
\right).
\end{align}
This zero mode is related to the translational invariance of the theory \eqref{NLsM1} that is explicitly broken by the skyrmion solution. The real and imaginary part of the amplitude of the eigenfunction \eqref{ZeroMode} correspond to  translations of the skyrmion within the two-dimensional plane. We have not yet found bound modes with a single or more nodes, which might however emerge for $m=-1$ at larger fields. 

\section{High-energy scattering of magnons}
\label{sec:HighEnergyScatt}

The properties of the magnon scattering states for arbitrary energies, $\varepsilon \geq \varepsilon_{\rm gap},$ have been discussed in Ref.~\onlinecite{Schuette2014}. In the present work, we elaborate on the scattering of magnons in the high-energy limit, $\varepsilon \gg \varepsilon_{\rm gap}$, which corresponds to magnon wavevectors much larger than the inverse skyrmion radius, $k r_s \gg 1$. In this limit, the treatment of the scattering simplifies considerably allowing for a transparent discussion of characteristic features. 

In the high-energy limit the magnon-skyrmion interaction is governed by the scattering vector potential $\vec a({\bf r}) = a^\chi(\rho) \hat \chi$ of Eq.~\eqref{VectorPot} so that the scattering has a purely magnetic character. 
In particular, in this limit one can neglect the anomalous potential $\mathcal{V}_x$, and the BdG equation \eqref{BdG} reduces to a Schr\"odinger equation for the magnon wavefunction
\begin{align} \label{Schroedinger}
i \hbar \partial_t \psi  
=\Big( \frac{\hbar^2 (- i \vec \nabla - \vec a)^2}{2 M_{\rm mag}}  + \varepsilon_{\rm gap}
\Big) \psi .
\end{align}
Setting $\psi({\bf r}, t) = e^{-i\varepsilon_k t/\hbar} e^{i m\chi} \eta_m(\rho)$ with the dispersion $\varepsilon_k = \varepsilon_{\rm gap} + \frac{\hbar^2 k^2}{2 M_{\rm mag}}$ and wavevector $k>0$, one obtains the radial wave equation for $\eta_m(\rho)$
\begin{align}
\Big[-  \Big(\partial_\rho^2 + \frac{\partial_\rho}{\rho}\Big) + \frac{(m - \rho a^\chi(\rho))^2}{\rho^2}  -k^2
\Big] \eta_m =0.
\end{align}
For large distances $\rho a^\chi(\rho) \to 1$, which identifies the angular momentum of the incoming wave to be $L_z = \hbar (m-1)$.

\subsection{Eikonal approximation}

\begin{figure}
\centering
\includegraphics[width=0.8\columnwidth]{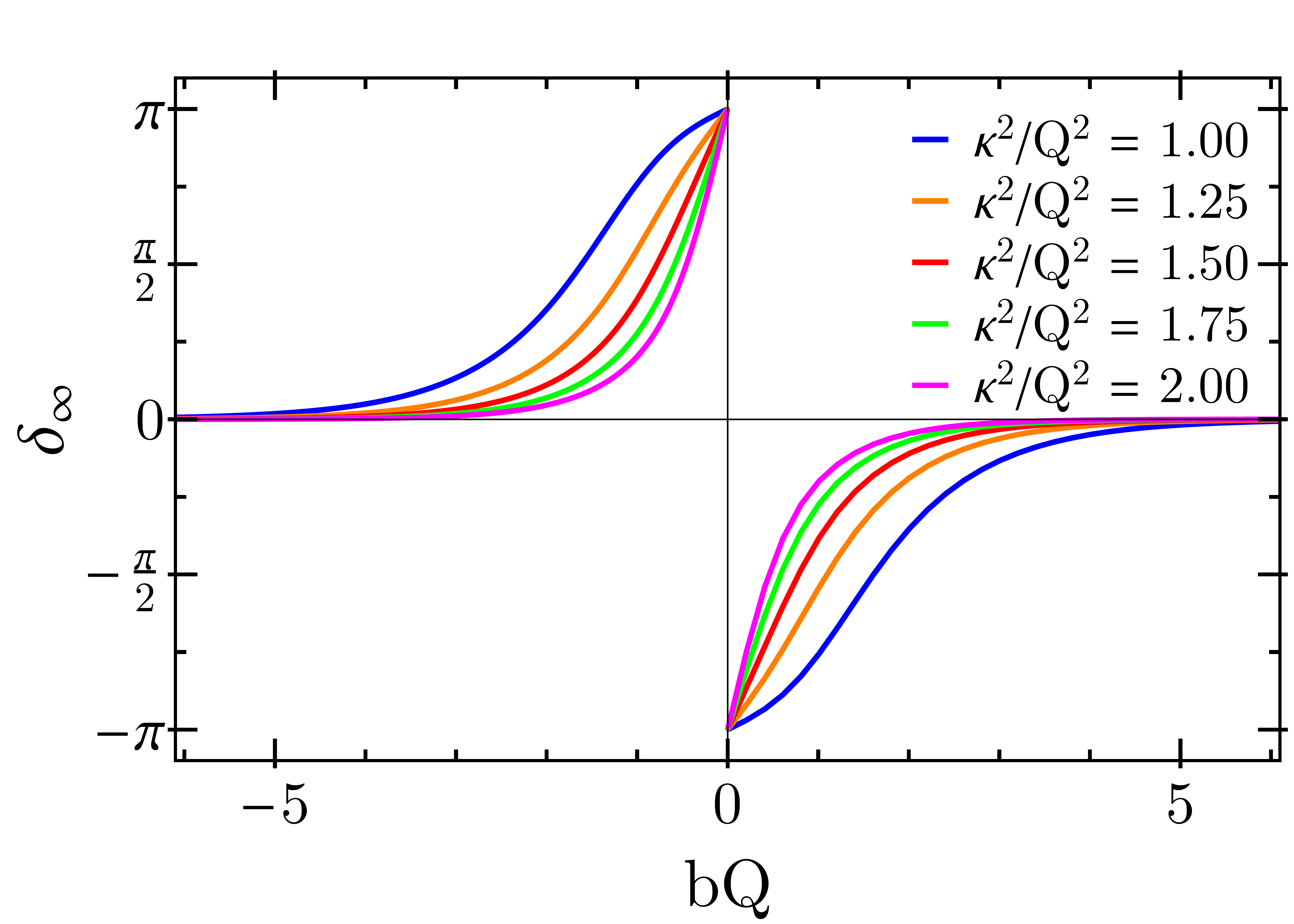}
\caption{Scattering phase shift for high-energy magnons \eqref{PSEikonal} as a function of impact parameter $b$ for different values of $\kappa^2/Q^2$. The scattering is non-perturbative as the phase shift assumes values within the entire interval $(-\pi,\pi)$. 
}
\label{fig:PhaseShift}
\end{figure}

As we are interested in the high-energy limit, we can treat this wave equation in the {\it eikonal approximation}. However, in order to make contact with Ref.~\onlinecite{Schuette2014}, we first give the resulting phase shift within the WKB approximation that is obtained by following Langer \cite{Langer1937,Berry1972}
\begin{align}
\delta^{\rm WKB}_m &= \int_{\rho_0}^\infty \Big( \sqrt{k^2 - \frac{(m - \rho a^\chi(\rho))^2}{\rho^2}} - k \Big) d\rho 
\nn
\\&+ \frac{\pi}{2} |m-1| - k \rho_0
\end{align}
where $\rho_0$ is the classical turning point. The eikonal approximation for the phase shift  is then obtained by taking the limit $k \to \infty$ while keeping the impact parameter $b = L_z/(\hbar k)$ fixed, $\delta^{\rm WKB}_m \to \delta_\infty(b)$, yielding 
\begin{align} \label{PSEikonal}
\delta_{\infty}(b) = b \int_{|b|}^\infty \frac{a^\chi_{\rm lab}(\rho)}{ \sqrt{\rho^2- b^2}} d\rho 
= b \int_{1}^\infty \frac{a^\chi_{\rm lab}(s |b|)}{\sqrt{s^2 - 1}} ds 
\end{align}
where we used $\rho a^\chi_{\rm lab}(\rho) = \rho a^\chi(\rho) -1$, see Eq.~\eqref{VectorPotLab}, and in the last equation we substituted $s = \rho/|b|$. This phase shift is odd with respect to $b$, i.e.~$\delta_{\infty}(b) = - \delta_{\infty}(-b)$.
Note that the scattering is non-perturbative even  in the high-energy limit 
in the sense that the phase shift $\delta_{\infty}(b)$ covers the entire interval $(-\pi,\pi)$ as a function of $b$, see Fig.~\ref{fig:PhaseShift}.
In particular, in the limit of small impact parameter $b \to 0$:
\begin{align} \label{PSstep}
\delta_{\infty}(b) \to b \int_{1}^\infty \frac{-2/(s |b|)}{\sqrt{s^2 - 1}} ds = - \pi\, {\rm sgn}(b).
\end{align}
For impact parameters larger than the skyrmion radius, $b \gg r_s$, the phase shift vanishes exponentially.

\begin{figure}
\centering
\includegraphics[width=0.9\columnwidth]{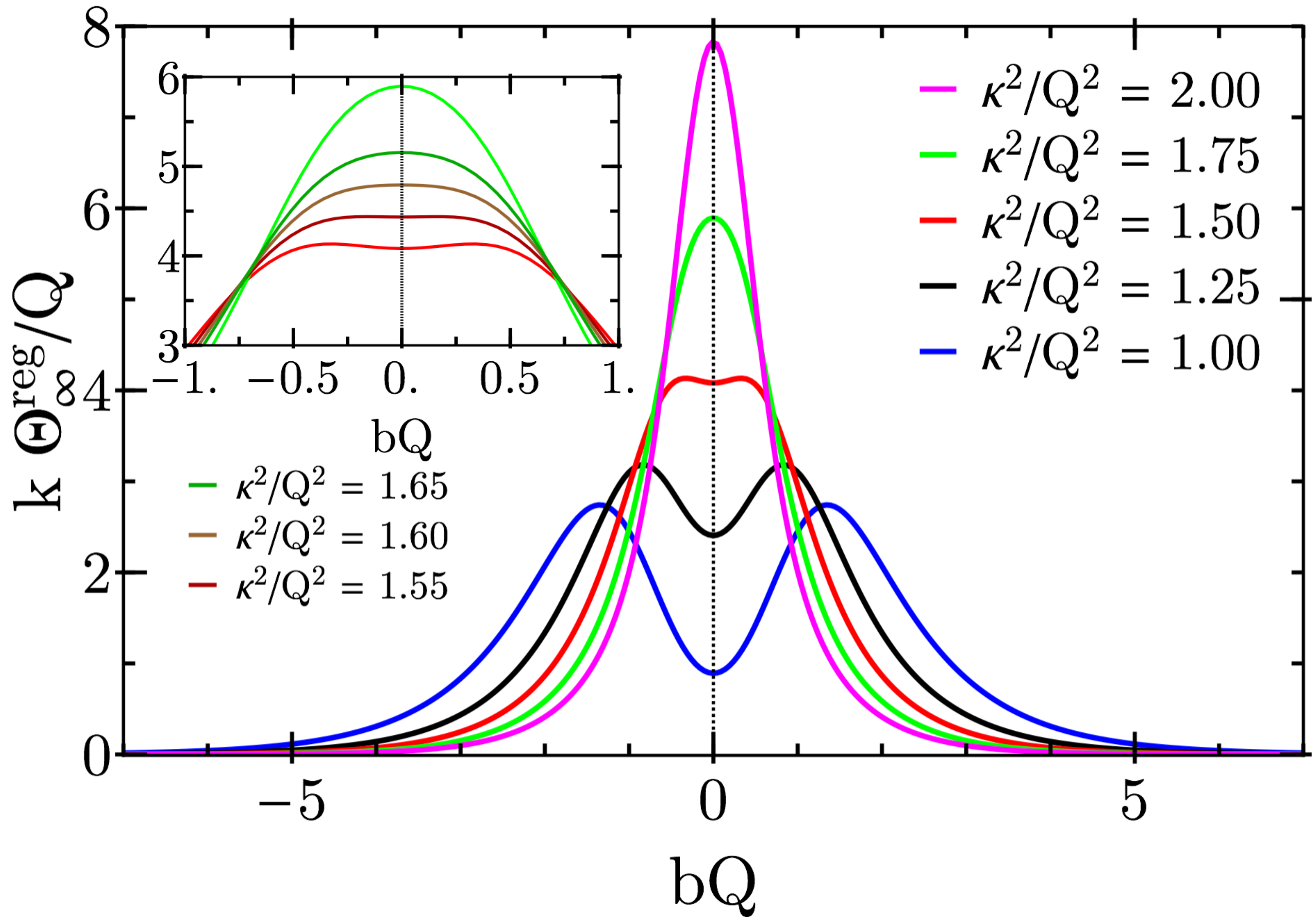}
\caption{Classical deflection angle for scattering of high-energy magnons \eqref{ClassicalDeflectionAngle} as a function of impact parameter $b$  for different values of $\kappa^2/Q^2$. 
In the high-energy limit, the scattering is in the forward direction with a deflection angle decreasing with increasing wavevector $k$ as $\Theta^{\rm reg}_{\infty}(b) \sim 1/k$. The inset focuses on the change of curvature at $b=0$ for $\kappa^2 \approx 1.6 Q^2$ with the same units on the vertical axis.}
\label{fig:DeflectionAngle}
\end{figure}

The deflection angle in the eikonal approximation is given by the derivative of $\delta_{\infty}(b)$,
\begin{align}
&\Theta_{\infty}(b) = 2 \hbar \frac{\partial \delta_\infty(b)}{\partial L_z} = \frac{2}{k}  \delta'_\infty(b) = \Theta^{\rm reg}_{\infty}(b) - \frac{4\pi}{k} \delta(b).
\end{align}
The step of $\delta_{\infty}(b)$ for head-on collisions, see Eq.~\eqref{PSstep}, leads to the delta function $\delta(b)$. The classical deflection function is given by the regular part, which reads 
\begin{align} \label{ClassicalDeflectionAngle}
\Theta^{\rm reg}_{\infty}(b)  
&= \frac{2}{\hbar k}  \int_{1}^\infty \frac{s |b| \mathcal{B}_{\rm reg}(s |b|)}{\sqrt{s^2 - 1}} ds
\\
&= \frac{1}{\hbar k}  \int_{-\infty}^\infty   \mathcal{B}_{\rm reg}\Big(\sqrt{b^2+x^2}\Big) dx,
\end{align}
where in the last equation we substituted  $x = |b| \sqrt{s^2 -1}$ and used that the integrand is an even function of $x$.
It is determined by the regular part of the flux density, $\mathcal{B}_{\rm reg}$, given in Eq.~\eqref{Breg}, integrated along a straight trajectory shifted from the $x$-axis by the impact parameter $b$.
Its behavior as a function of $b$ is shown in Fig.~\ref{fig:DeflectionAngle} for various values of $\kappa^2/Q^2$. 
The deflection angle is always positive implying that, classically, the Lorentz force attributed to $\mathcal{B}_{\rm reg}$ always skew scatters the magnons to the right-hand side from the perspective of the incoming wave even for negative impact parameters, see Fig.~\ref{fig:Illustration}(b). 
Note that the deflection angle possesses a local minimum at $b=0$ for $\kappa^2 \lesssim 1.6 Q^2$, 
that however gets filled and transitions into a maximum for larger values of $\kappa$. This change of curvature at $b=0$ is related to the change of curvature of the flux density $\mathcal{B}''_{\rm reg}(\rho)$ at the origin $\rho=0$, see Fig.~\ref{fig:Flux}, that happens for a similar value of $\kappa$.
As the total flux of $\mathcal{B}_{\rm reg}$ is quantized, the deflection angle integrated over the impact parameter is just given by the universal value $\int_{-\infty}^\infty db\, \Theta^{\rm reg}_{\infty}(b) = 4\pi/k$.

\subsection{Differential cross section}

In the following, we consider a magnon scattering setup where an on-shell magnon plane wave with wavevector ${\bf k} = k \hat x$ along the $x$-direction and amplitude $A$ defined within the laboratory orthogonal frame, see Eq.~\eqref{GaugeTransformation}, is impinging on the skyrmion, see also Fig.~\ref{fig:Illustration}(b). At large distances this wavefunction assumes the asymptotic behavior
\begin{align} \label{ScatteringWavefunction}
\psi_{\rm lab}({\bf r},t) = A e^{- i\varepsilon_k/\hbar} \left(e^{i {\bf k} {\bf r}} + f(\chi) \frac{e^{i k \rho}}{\sqrt{\rho}}\right),
\end{align}
where the scattering amplitude is given by 
\begin{align} \label{ScattAmplitudeExact}
f(\chi) = \frac{e^{-i \pi/4}}{\sqrt{2\pi k}} \sum^\infty_{m=-\infty}\, e^{i (m-1) \chi} (e^{i 2 \delta_{m}} - 1).
\end{align}
Note that the additional phase factor $e^{- i \chi}$ arises from the gauge transformation \eqref{GaugeTransformation}. The differential cross section is then obtained by $\frac{\partial \sigma}{\partial \chi} = |f(\chi)|^2$.

\subsubsection{High-energy limit of the scattering amplitude}

\begin{figure}
\centering
\includegraphics[width=0.9\columnwidth]{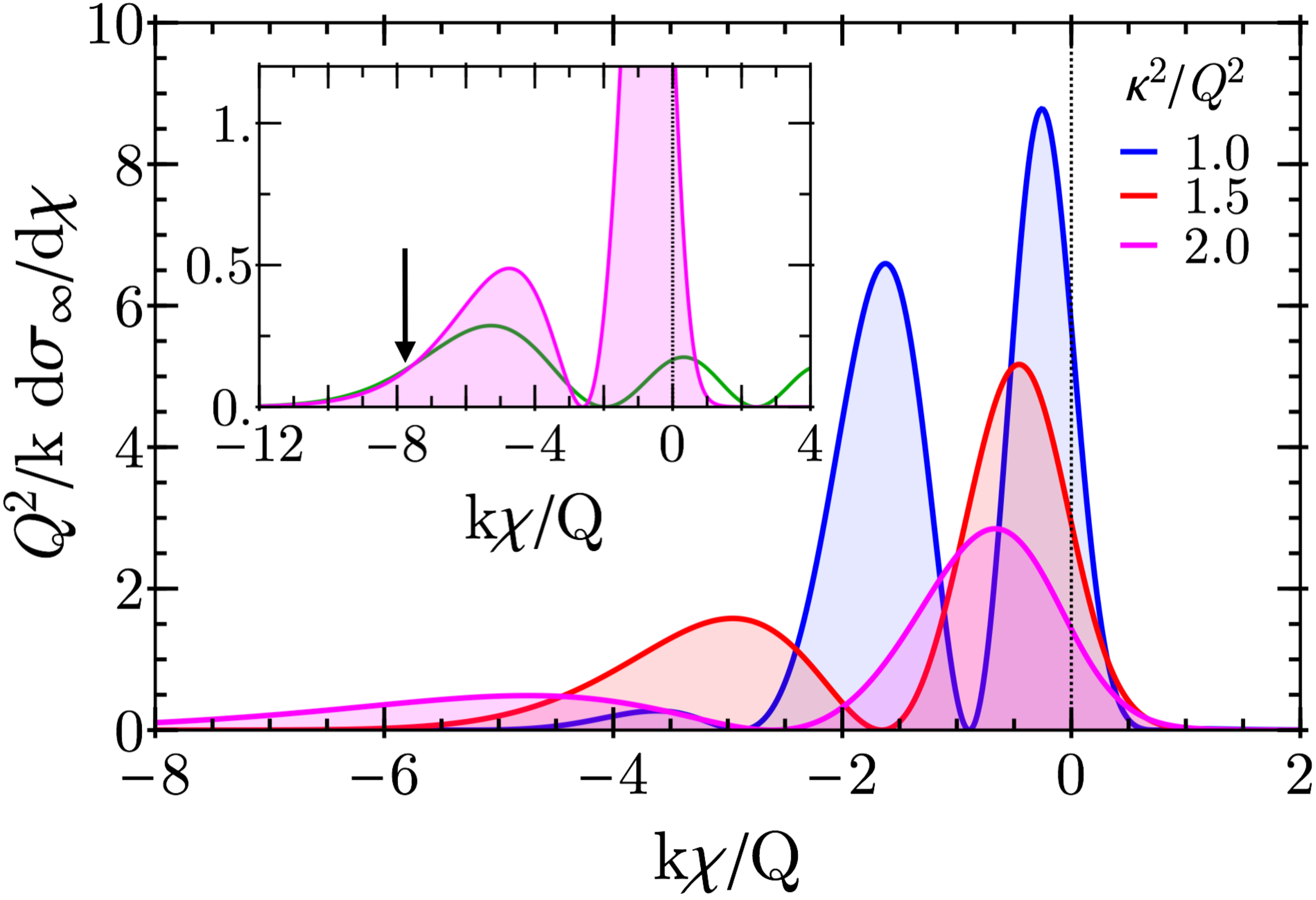}
\caption{Differential cross section of high-energy magnons \eqref{DiffCross} for various values of $\kappa^2/Q^2$. It is asymmetric with respect to $\chi = 0$ due to skew scattering, and the oscillations are attributed to rainbow scattering. The inset compares the curve for $\kappa^2/Q^2 = 2$ with the Airy approximation \eqref{ScattAmplitudeAiry} (green solid line) with the same units on the vertical axis; the arrow indicates the position of the corresponding rainbow angle $-k \Theta_\infty^{\rm reg}(0)/Q$.
}
\label{fig:DiffCross}
\end{figure}

In the high-energy limit, we can replace the sum over angular momentum numbers by an integral over the impact parameter, $b = (m-1)/k$,  so that the scattering amplitude reads approximately
\begin{align} \label{ScattAmplitude}
f_{\infty}(\chi) = \frac{e^{-i \pi/4}}{\sqrt{2\pi k}} k \int_{-\infty}^\infty db\, e^{i b k \chi} (e^{i 2 \delta_\infty(b)} - 1),
\end{align}
with $\delta_\infty(b)$ defined in Eq.~\eqref{PSEikonal}. The differential cross section in this limit,
\begin{align} \label{DiffCross}
\frac{\partial \sigma_\infty}{\partial \chi} = |f_\infty(\chi)|^2 = \frac{k}{Q^2}\, S\Big(k \chi/Q\Big),
\end{align}
is then determined by the dimensionless function $S$, which is shown in Fig.~\ref{fig:DiffCross}.
 
The support of the differential cross section is approximately limited by the extremal values of the classical deflection angle of Eq.~\eqref{ClassicalDeflectionAngle} and Fig.~\ref{fig:DeflectionAngle}. Note that the angle $\chi$ is defined in a mathematically positive sense so that a positive $\Theta$ translates to a negative value of $\chi$. It is strongly asymmetric with respect to forward scattering reflecting the {\it skew scattering} arising from the Lorentz force of the emerging magnetic field $\mathcal{B}_{\rm reg}$. 
 
\subsubsection{Rainbow scattering and Airy approximation}

Moreover, the differential cross section exhibits oscillations. These can be attributed to an effect known as {\it rainbow scattering}. As the function $\Theta_\infty^{\rm reg}(b)$ is even in $b$, there exist for a given classically allowed deflection angle $\Theta$ always at least one pair $\pm b_{\rm cl}$ of impact parameters that solve $\Theta_\infty^{\rm reg}(\pm b_{\rm cl}) = \Theta$. For a given angle $\Theta$ the magnons might, therefore, either pass the skyrmion on its right- or left-hand side; these classical trajectories interfere leading to the oscillations in $d\sigma/d\chi$. 

First, consider values $\kappa^2 \gtrsim 1.6 Q^2$ for which $\Theta_\infty^{\rm reg}(b)$ possesses only a single maximum at $b=0$. The maximum value $\Theta_\infty^{\rm reg}(0)$ is  known as rainbow angle and for values of $\chi$ close to $- \Theta_\infty^{\rm reg}(0)$, the interference effect of classical trajectories can be illustrated with the help of the Airy approximation for the scattering amplitude. For such values of $\chi$, the $-1$ in the integrand of Eq.~\eqref{ScattAmplitude} can be neglected as it only contributes to forward scattering. Expanding the exponent of the remaining integrand up to third order in $b$ one then obtains
\begin{align} \label{ScattAmplitudeAiry}
&f_{\infty}(\chi)\Big|_{\rm Airy} = \\\nn
&= \frac{e^{-i \pi/4}}{\sqrt{2\pi k}} k \int\limits_{-\infty}^\infty db\, 
\exp\Big[i b k (\chi + \Theta_\infty^{\rm reg}(0)) + i  \frac{k}{6} {\Theta''}^{\rm reg}_\infty(0) b^3 \Big]
\\\nn
& = \frac{\sqrt{2\pi k}\, e^{-i \pi/4}}{[k |{\Theta''}^{\rm reg}_\infty(0)|/2]^{1/3}} {\rm Ai}\Big(- \frac{k (\chi + \Theta_\infty^{\rm reg}(0))}{[k |{\Theta''}^{\rm reg}_\infty(0)|/2]^{1/3}}\Big),
\end{align}
where in the last equation we identified the integral representation of the Airy function Ai using that ${\Theta''}^{\rm reg}_\infty(0) < 0$.
 
In the inset of Fig.~\ref{fig:DiffCross}, we compare the differential cross section at $\kappa^2 = 2 Q^2$ with the Airy approximation resulting from Eq.~\eqref{ScattAmplitudeAiry}. The latter reproduces the exponential decrease for large angles $\chi < -\Theta_\infty^{\rm reg}(0)$ corresponding to the dark side and also the oscillations on the bright side, $\chi > -\Theta_\infty^{\rm reg}(0)$, of the rainbow angle. It of course fails close to forward scattering and for positive angles $\chi > 0$ where the classical deflection angle has lost its support.

Close to $\kappa^2 \approx 1.6 Q^2$ even the derivative ${\Theta''}^{\rm reg}_\infty(0)$ vanishes, see inset of Fig.~\ref{fig:DeflectionAngle}, giving rise to a {\it cubic rainbow} effect.\cite{Connor1970} Finally, for smaller values of $\kappa^2$ there also exist two pairs of classical trajectories that interfere in the differential cross section.


\subsection{Total and transport scattering cross section}

 \begin{figure}
\centering
\includegraphics[width=0.8\columnwidth]{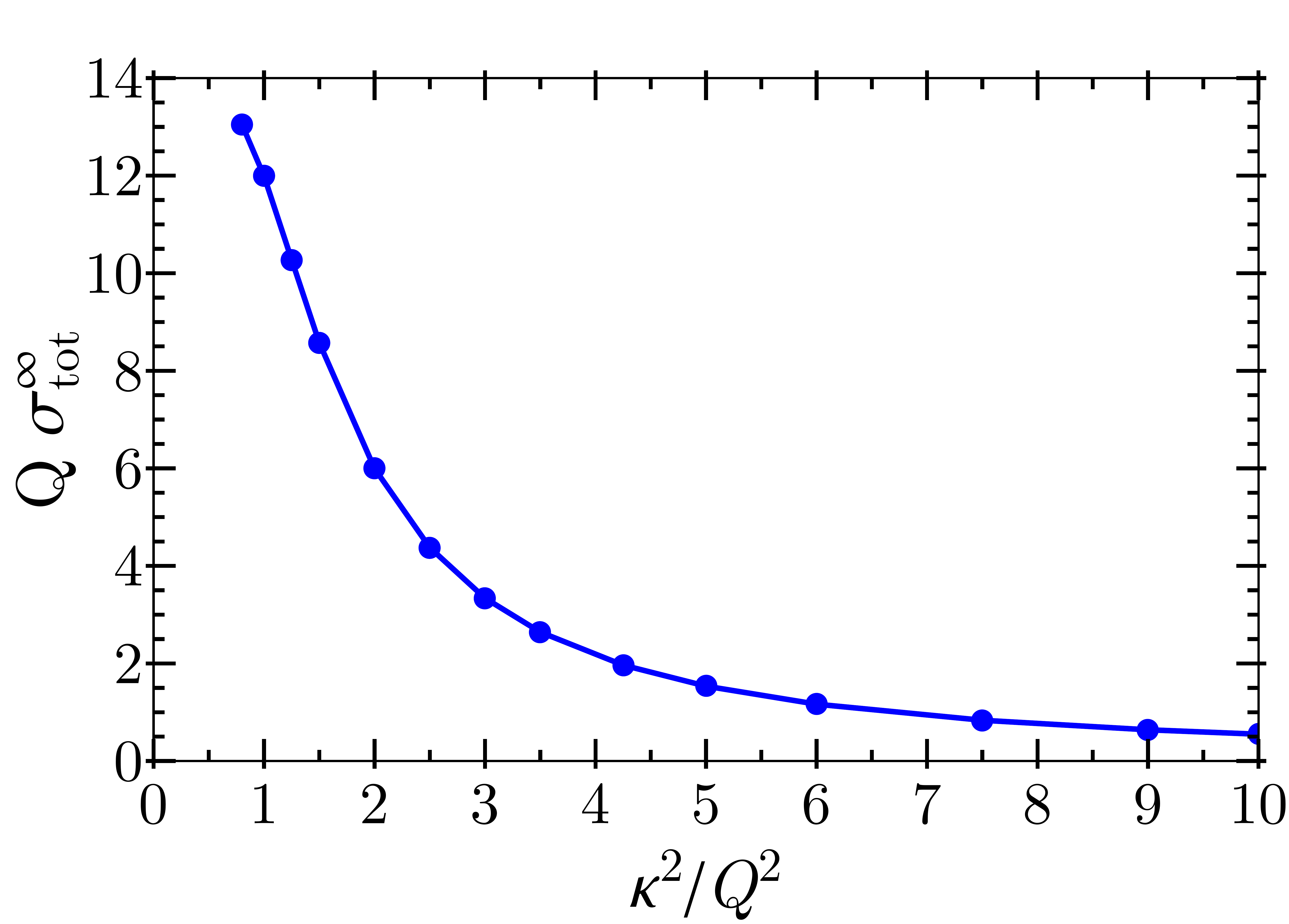}
\caption{Total scattering cross section of the skyrmion in the high-energy limit, Eq.~\eqref{SigmaTot}, as a function of $\kappa^2/Q^2$. 
It decreases for increasing external magnetic field strength $\kappa^2$.
}
\label{fig:sigmatot}
\end{figure}

We continue with a discussion of the total, $\sigma_{\rm tot} = \int_{-\pi}^\pi d\chi d\sigma/d\chi$, and the transport scattering cross section defined in Eq.~\eqref{TrCS}. In order to determine their high-energy limit, one first expresses $d\sigma/d\chi = |f(\chi)|^2$ in terms of the exact representation \eqref{ScattAmplitudeExact} for the scattering amplitude $f(\chi)$ and evaluates the integral over $\chi$. 
Afterwards one takes the high-energy limit $k \to \infty$ with keeping the impact parameter $b = (m-1)/k$ fixed. 

The total scattering cross section of the skyrmion then reduces to
\begin{align} \label{SigmaTot}
\sigma^{\infty}_{\rm tot} = 4 \int_{-\infty}^\infty db\, (\sin \delta_{\infty}(b))^2 .
\end{align}
It saturates to a finite value in the high-energy limit, and its dependence on $\kappa$ is shown in Fig.~\ref{fig:sigmatot}. It decreases with increasing $\kappa$ and thus decreasing skyrmion radius $r_s$ as expected. One might expect that $\sigma_{\rm tot}^\infty \sim r_s$ which however only holds approximately. 

Using that $\delta_\infty(b)$ is an odd function of $b$, we obtain for the transport scattering cross section $\sigma_\perp(\varepsilon)$ in the high-energy limit
\begin{align} \label{TSCSperp}
\sigma^\infty_\perp(\varepsilon) = \frac{8}{k} \int_{0}^\infty db\, \delta'_\infty(b) (\sin \delta_\infty(b))^2 =
\\ 
=\frac{8}{k} \Big[\frac{\delta_\infty}{2} - \frac{\sin(2\delta_\infty)}{4}\Big]^0_{-\pi} = \frac{4\pi}{k}.
\end{align}
In the last line, we further used the boundary values of the function $\delta_\infty(b)$. It vanishes $\sigma^\infty_\perp(\varepsilon) \sim 1/k$, but with a universal prefactor that is independent of $\kappa$.

\begin{figure}
\centering
\includegraphics[width=0.8\columnwidth]{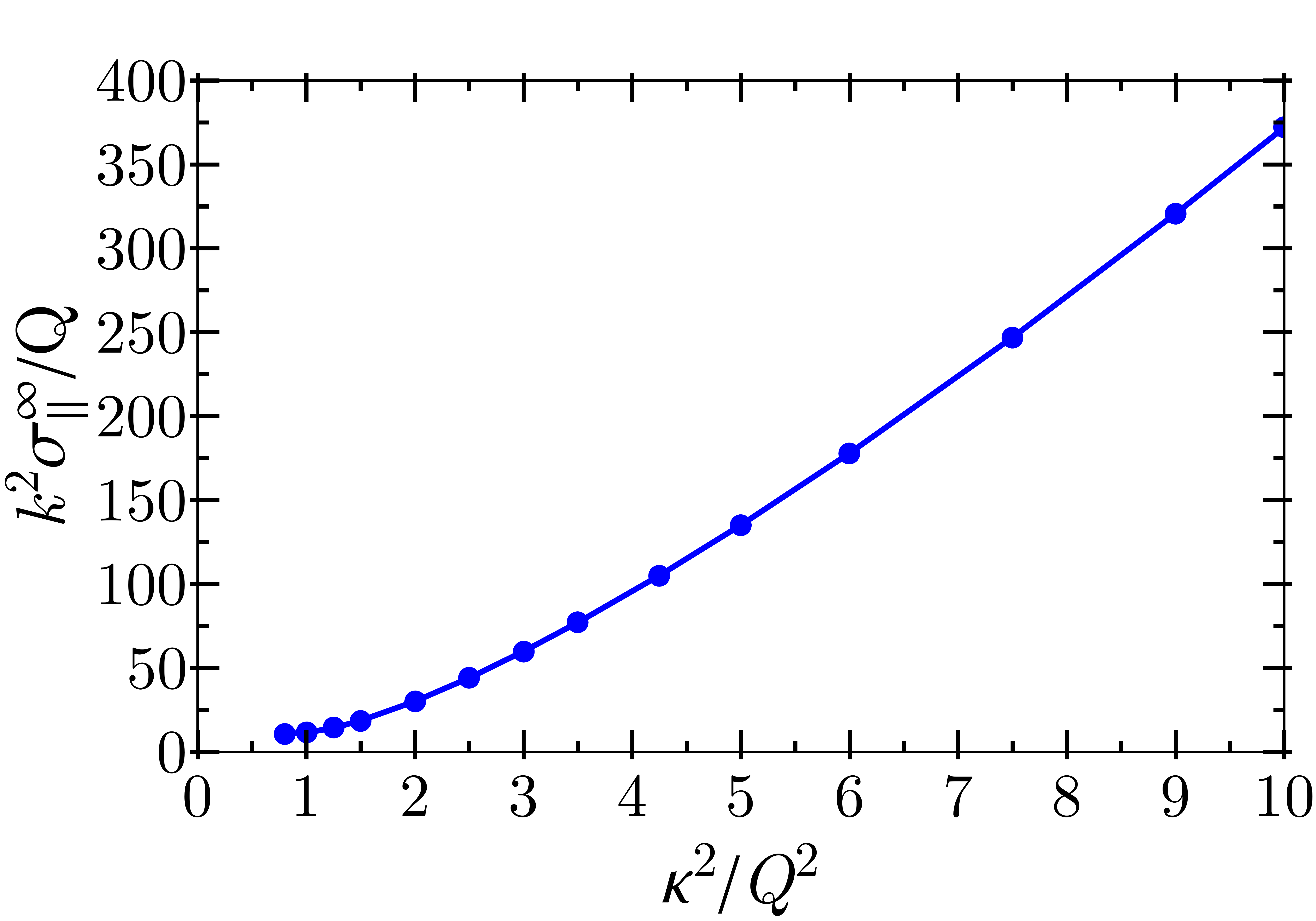}
\caption{The longitudinal transport scattering cross section, Eq.~\eqref{SigmaPara}, vanishes as $\sigma^\infty_\parallel \sim 1/k^2$ in the high-energy limit. The panel shows the $\kappa$-dependence of the prefactor.
}
\label{fig:sigmapara}
\end{figure}

Finally, for the ongitudinal transport scattering cross section we obtain for $k r_s \gg 1$
\begin{align}
\sigma^\infty_\parallel(\varepsilon) = \frac{4}{k^2} \int \limits_{0}^\infty db\, 
 \Big(2 (\delta'_\infty)^2 (\sin \delta_\infty)^2 - \delta''_\infty \sin \delta_\infty \cos \delta_\infty\Big).
\end{align}
After integrating by parts this simplifies to 
\begin{align} \label{SigmaPara}
\sigma^\infty_\parallel(\varepsilon) = \frac{4}{k^2} \int_{0}^{\infty} db\, (\delta'_\infty(b))^2 = \int_{-\infty}^{\infty} db\, \frac{1}{2}(\Theta^{\rm reg}_\infty(b))^2.
\end{align}
It is given by the square of the classical deflection angle \eqref{ClassicalDeflectionAngle} integrated over the impact parameter $b$. It vanishes as $\sigma^\infty_\parallel \sim 1/k^2$ in the high-energy limit with a prefactor whose $\kappa$ dependence is shown in Fig.~\ref{fig:sigmapara}. On dimensional grounds one might expect $k^2 \sigma_\parallel^\infty \sim 1/r_s$, which again only 
holds approximately. 

\subsection{Magnon pressure in the high-energy limit}

We have shown in Ref.~\onlinecite{Schuette2014} by considering the energy-momentum tensor of the field theory that the monochromatic plane wave of \eqref{ScatteringWavefunction} with wavevector ${\bf k} = k \hat x$ leads to a momentum-transfer force in the Thiele equation of motion of the form given in Eq.~\eqref{MomentumTransferThiele} with the magnon current 
\begin{align}
\vec J_\varepsilon = \hat x |A|^2 \frac{m_0\hbar}{g \mu_B} \frac{\hbar k}{M_{\rm mag}} = \frac{|\vec G|}{4\pi}\vec v_{\rm eff}.
\end{align}
In the second equation, we have introduced the effective velocity $\vec v_{\rm eff} = \hat x  |A|^2 \frac{\hbar k}{M_{\rm mag}}$ and $|\vec G| = 4\pi m_0\hbar/(g\mu_B)$ with the purpose of comparing with Eq.~\eqref{AdiabaticThiele}. 

\begin{figure}
\centering
\includegraphics[width=0.8\columnwidth]{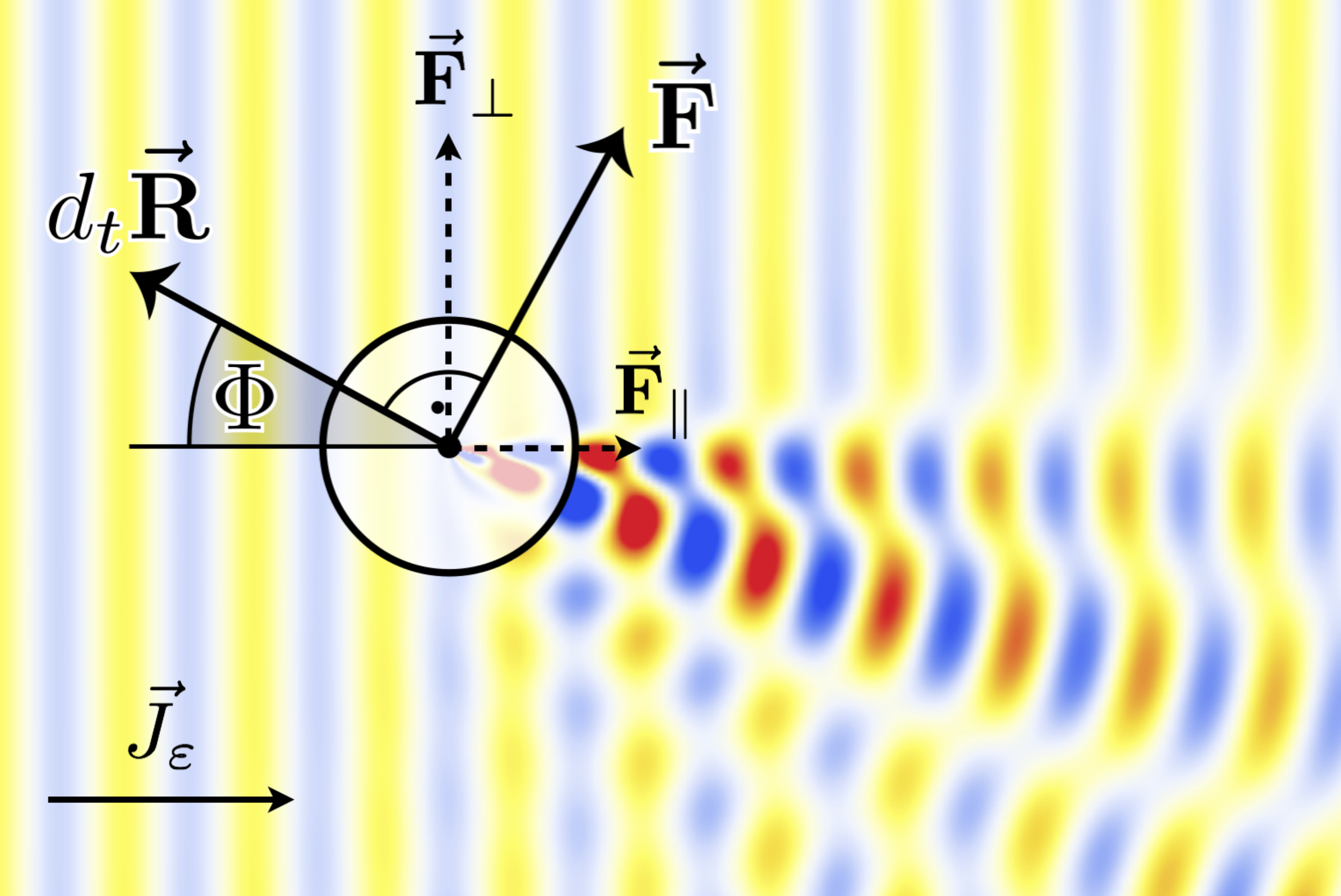}
\caption{An incoming monochromatic magnon current $\vec J_\varepsilon$ leads to a momentum-transfer force $\vec F$ that is determined by the transport scattering cross sections, see Eq.~\eqref{MomentumTransferThiele}. The image shows the magnon wavefunction in the WKB approximation with the skyrmion being represented by the circle with radius $r_s$ [29]. For high-energy magnons with wavevector $k r_s \gg 1$, the transversal force dominates, $F_\parallel /F_\perp \sim 1/k$, resulting in a skyrmion motion $\partial_t \vec R$ approximately antiparallel to $\vec J_\varepsilon$ with a small skyrmion Hall angle $\Phi \sim 1/k$.
}
\label{fig:pressure}
\end{figure}

This momentum transfer is illustrated in Fig.~\ref{fig:pressure}. In the high-energy limit, the transversal and longitudinal forces are given by
\begin{align} \label{Fperp}
\vec F_{\perp} & = k \sigma^{\infty}_\perp(\varepsilon) (\hat z \times \vec J_\varepsilon) = 4\pi (\hat z \times \vec J_\varepsilon) = - \vec G \times \vec v_{\rm eff},
\\ \label{Fpara}
\vec F_{\parallel} &= 
k \sigma^\infty_\parallel(\varepsilon)
\vec J_\varepsilon
= 
\frac{|\vec G|}{8\pi} k
\int_{-\infty}^{\infty} db\, (\Theta^{\rm reg}_\infty(b))^2 \vec v_{\rm eff},
\end{align}
where we used Eqs.~\eqref{TSCSperp} and \eqref{SigmaPara} as well as $\vec G = - |\vec G| \hat z$.
They are indeed of the form given in Eq.~\eqref{AdiabaticThiele}. The transversal momentum-transfer force, $\vec F_{\perp}$, is universal, and $\vec F_{\parallel}$ is determined by the $\beta$ parameter of Eq.~\eqref{BetaReactive} after identifying $\Theta(b)$ with the classical deflection angle $\Theta^{\rm reg}_\infty(b)$.

Is there an intuitive classical interpretation of these momentum-transfer forces? From the classical limit of the Schr\"odinger equation \eqref{Schroedinger} follows the equation of motion for the coordinate $\vec r(t)$ of a classical magnon particle\cite{Mochizuki2014}
\begin{align} \label{EoM}
M_{\rm mag} \ddot{\vec r} = \dot{\vec r} \times (\hat z \mathcal{B}_{\rm reg}(|\vec r|)),
\end{align}
with the regular part of the effective magnetic flux distribution $\mathcal{B}_{\rm reg}$ of Eq.~\eqref{Breg}.
Note that we have chosen in Eq.~\eqref{Schroedinger} the charge to be $+1$.
Consider the change of momentum, $\delta \vec p$, of this magnon particle after scattering off  the static skyrmion by integrating the left-hand side of Eq.~\eqref{EoM},
\begin{align} \label{MomentumChange}
\delta \vec p(b) &= \int^\infty_{-\infty} dt\, M_{\rm mag} \ddot{\vec r}(t) =
 M_{\rm mag} (\dot{\vec r}(\infty) - \dot{\vec r}(-\infty)) 
 \nn\\&= p 
 \left(\begin{array}{c} \cos \Theta(b) - 1 \\ - \sin \Theta(b) \end{array}\right) .
\end{align}
In the last equation, we have exploited that at large distances the magnitude of momentum $M_{\rm mag} |\dot{\vec r}(\pm\infty)| = p$ remains unchanged due to energy conservation, while the orientation of velocity is determined by the scattering angle $\Theta(b)$, see Fig.~\ref{fig:Illustration}(b), that depends on the impact parameter $b$ of the trajectory. 

This momentum $\delta \vec p(b)$ is transferred to the skyrmion. The momentum-transfer force on the skyrmion due to a current of classical magnon particles along $\hat x$ with density $m_0/(g\mu_B)$ and velocity $v_{\rm eff} = |\vec v_{\rm eff}|$ is then given by
\begin{align}
\vec F = \left(\begin{array}{c} F_\parallel \\ F_\perp \end{array}\right) = - v_{\rm eff} \frac{m_0}{g\mu_B} \int_{-\infty}^\infty db\, \delta \vec p(b),
\end{align}
with $F_{\parallel/\perp} = |\vec F_{\parallel/\perp}|$. In the high-energy limit, the scattering is in forward direction so that we can expand Eq.~\eqref{MomentumChange} in the deflection angle $\Theta(b)$ and the force becomes with $p = \hbar k$
\begin{align}
\vec F = v_{\rm eff} \frac{m_0}{g\mu_B} \hbar k \int_{-\infty}^\infty db 
\left(\begin{array}{c} \frac{1}{2} (\Theta(b))^2 \\ \Theta(b) \end{array}\right).
\end{align}
Finally using that the integral $\int_{-\infty}^\infty db \Theta(b) = 4\pi/k$ is quantized in the high-energy limit, that we already know from the discussion in the context of Eq.~\eqref{ClassicalDeflectionAngle}, we recover Eqs.~\eqref{Fperp} and \eqref{Fpara}.

For the understanding of the universality of $F_\perp$, it is also instructive 
to consider alternatively the right-hand side of the classical equations of motion \eqref{EoM}. By integrating the right-hand side, one obtains for the transversal momentum change
\begin{align}
\delta p_y = \int_{-\infty}^\infty dt (- \dot{x}) \mathcal{B}_{\rm reg}(|\vec r|) \approx - \int_{-\infty}^{\infty} dx \mathcal{B}_{\rm reg}(\sqrt{b^2 + x^2}).
\end{align}
In the last equation we employed the high-energy approximation by straightening the magnon trajectory.
It follows then for the transversal force
\begin{align}
F_\perp &= v_{\rm eff} \frac{m_0}{g\mu_B} \int_{-\infty}^{\infty} db \int_{-\infty}^{\infty} dx \mathcal{B}_{\rm reg}(\sqrt{b^2 + x^2}) 
\\&= v_{\rm eff} \frac{m_0}{g\mu_B} 4\pi \hbar,
\end{align}
where its universality is now directly related to the quantized total flux of $\mathcal{B}_{\rm reg}$.

\section{Summary}
\label{sec:discussion}

The scattering of high-energy magnons with wavevectors $k r_s \gg 1$ off a magnetic skyrmion of linear size $r_s$ is governed by a vector scattering potential. The associated effective magnetic field 
is related to the topological charge density of the skyrmion and is exponentially confined to the skyrmion area. 
The total flux is determined by the topological skyrmion number and is quantized.

When a magnon traverses the skyrmion, classically speaking, it experiences the resulting Lorentz force 
and is deflected to a preferred direction determined by the sign of the emergent magnetic flux. This results in skew scattering with a differential cross section that is asymmetric with respect to forward scattering, see Fig.~\ref{fig:DiffCross}.
As the flux distribution is rotationally symmetric, the classical deflection angle $\Theta(b)$ as a function of the impact parameter $b$ is even in the high-energy limit, $\Theta(b) = \Theta(-b)$. 
As a consequence, for a given deflection angle $\Theta$ there exist corresponding classical trajectories with positive as well as negative $b$, i.e., that pass the skyrmion on the left-hand as well as on the right-hand side. These trajectories interfere which leads to oscillations in the differential cross section, an effect known as rainbow scattering. 

Magnons hitting the skyrmion also transfer momentum giving rise to a force in the Thiele equation of motion, see Eq.~\eqref{MomentumTransferThiele}. In the high-energy limit, this force can be interpreted classically and assumes the form of Eq.~\eqref{AdiabaticThiele}. While the transversal momentum-transfer force, $F_\perp$ is universal and determined by the total emergent magnetic flux, the longitudinal momentum-transfer force, $F_\parallel$ is obtained by integrating $(\Theta(b))^2$ over the impact parameter $b$ leading to the parameter $\beta_\varepsilon$ of Eq.~\eqref{BetaReactive}. Since for large energies the classical deflection angle is small, $\Theta(b) \sim 1/k$, the momentum transfer is mainly transversal, $F_\parallel/F_{\perp} \sim 1/k$. This leads to a skyrmion motion $\partial_t \vec R$ approximately antiparallel to the magnon current $\vec J_\varepsilon$ with a small skyrmion Hall angle $\Phi = \beta_\varepsilon/|\vec G|$ defined in Fig.~\ref{fig:pressure}, 
\begin{align} \label{Hall}
\Phi = \frac{1}{2} \frac{\int_{-\infty}^\infty (\Theta(b))^2 db}{\int_{-\infty}^\infty \Theta(b) db} = 
\frac{k}{8\pi} \int_{-\infty}^\infty (\Theta(b))^2 db \propto \frac{1}{k},
\end{align}
where  the integral $\int_{-\infty}^\infty \Theta(b) db = 4\pi/k$ is universal 
in the high-energy limit.
Interestingly, the Hall angle $\Phi$ at high energies increases with increasing $\kappa$, which is shown in Fig.~\ref{fig:sigmapara} identifying $\Phi = k \sigma_\parallel^\infty(\varepsilon)/4\pi$.

While the skyrmion Hall angle $\Phi$ is small at high energies $k r_s \gg 1$, we note that it increases with decreasing energy and assumes the maximum value\cite{Schuette2014} $\Phi = \pi/2$ in the low-energy limit $k r_s \ll 1$ where $s$-wave scattering prevails and Eq.~\eqref{AdiabaticThiele} ceases to be valid. 

 \acknowledgments

We acknowledge helpful discussions with M.~Mostovoy, A.~Rosch, and C. Sch\"utte. 


\end{document}